\documentclass[twocolumn,english,pra,floatfix,superscriptaddress]{revtex4-2}
\usepackage[T1]{fontenc}
\usepackage[latin9]{inputenc}
\usepackage{amstext}
\usepackage{amssymb,amsmath}
\usepackage{graphicx}
\usepackage{esint}
 \usepackage{latexsym}
\usepackage{xcolor}
\usepackage{braket}
\usepackage{soul}

\definecolor{lblue}{RGB}{51,71,158}
\usepackage[colorlinks=true,citecolor=blue,linkcolor=blue,urlcolor=lblue]{hyperref}

\makeatletter
\newenvironment{lyxlist}[1]
	{\begin{list}{}
		{\settowidth{\labelwidth}{#1}
		 \setlength{\leftmargin}{\labelwidth}
		 \addtolength{\leftmargin}{\labelsep}
		 }}
	{\end{list}}

\makeatother

\usepackage{babel}
\begin{document}
\title{Random Kronig-Penney-type potentials for ultracold atoms using dark states}
\author{Mateusz \L \k{a}cki}
\affiliation{Instytut Fizyki Teoretycznej, Wydzia\l{} Fizyki, Astronomii i Informatyki Stosowanej, Uniwersytet Jagiello\'{n}ski, ulica \L ojasiewicza 11, PL-30-348 Krak\'ow, Poland}
\email{mateusz.lacki@uj.edu.pl}
\author{Jakub Zakrzewski}
\affiliation{Instytut Fizyki Teoretycznej, Wydzia\l{} Fizyki, Astronomii i Informatyki Stosowanej, Uniwersytet Jagiello\'{n}ski, ulica \L ojasiewicza
11, PL-30-348 Krak\'ow, Poland}
\affiliation{Mark Kac Complex Systems Research Center, Uniwersytet Jagiello\'{n}ski, 
PL-30-348 Krak\'ow, Poland}
\email{jakub.zakrzewski@uj.edu.pl}

\begin{abstract}
A construction of a quasi-random potential for cold atoms using  dark states 
emerging in $\Lambda$ {level configuration}  is proposed. Speckle laser fields are used as a source
of randomness. 
 Anderson localisation 
in such potentials is studied  and compared  with the known results for the speckle
potential itself. It is found out that the localisation length is greatly decreased due to the non-linear
fashion in which dark-state potential is obtained. In effect, random dark state potentials  resemble those occurring in random Kronig-Penney-type Hamiltonians.
\end{abstract}

\maketitle

\section{Introduction}

A particle moving in the potential consisting
of narrow peaks may be  described by Kronig-Penney-type Hamiltonians \cite{Kronig1931}.
When the potential is periodic, the problem is solved by a simple Bloch approach.
The presence of disorder enriches the physics. Here one can imagine that periodicity is broken either by different potential amplitudes at periodically distributed sites - the case sometimes called a compositional disorder \cite{Izrailev2001} or by random position of scatterers having then structural (or positional) disorder. In both cases one typically expects Anderson localization \citep{Anderson1958} at all energies for one-dimensional (1D) system and uncorrelated disorder. The presence of correlations leads to mobility edges as predicted and verified experimentally for a number of models 
\citep{Soukoulis82,Flores89,Dunlap90,Diez94,Bellani99,Izrailev1999,Izrailev2001,Biddle10,Ganeshan15,Major18,Kohlert19}.

A standard way to implement potentials for ultracold atoms is to use
off-resonant laser standing waves via an AC Stark effect \citep{Jaksch1998}.
Such light-shift potentials enabled 
experiments typical for  condensed matter systems as manifested by e.g. the pioneering observation of Mott insulator
to superfluid quantum phase transition \citep{Greiner2002}. Later research {in optical lattice potentials} involved
the use of different atomic species that feature strong, long range interactions
\cite{Lahaye2009,Lewenstein2012,Dutta2015}, creation of  
 topological insulators \citep{Cooper2019} or studies of  non-equilibrium dynamics
\citep{Langen2015}. In particular, the 1D experiments with ultracold atoms in random potentials have been
conducted with the far off-resonant speckle potential \citep{Billy2008},
bichromatic fields \citep{Fallani2007} or digital mirror devices
\citep{Gauthier2016}. 

The AC Stark based approach leads, naturally, to  diffraction-limitations that prohibit creating
potentials with features much sharper than half of the laser wavelength.
To remedy that, a construction based on ultracold atoms in many-levels
coupling schemes \citep{Lacki2016,Jendrzejewski2016} was proposed.
Coherent population of a dark state in the three-levels $\Lambda$ configuration was used to create a periodic comb potential consisting of
subwavelength peaks \citep{Wang2018}. Involving more than three atomic levels \citep{Gvozdiovas2021,Kubala2021}
opens possibilities for more complex potentials \citep{Ruseckas2005,dalibard2011colloquium}. 

In this work we shall use a similar $\Lambda$  scheme to create random correlated potentials featuring sharp peaks, even beyond the diffraction limit. The underlying model and creation of a random dark state potential is described in Section \ref{sec:model}. The shapes of the potential peaks and basic statistical properties of the potential are quantitatively analyzed in Section~\ref{sec:dark-state-potential}. We focus on two cases. In Subsection~\ref{subsec:The-dark-state-lambda} both lasers forming the legs of the $\Lambda$ system are due to a speckle field. In Subsection~\ref{subsec:The-dark-state-running-lambda} one of them is due to a running wave, with corresponding potential consisting of equidistant sharp tall peaks of a (quasi)random height.

In Section~\ref{sec:localization}, we study Anderson localization of a random potential from Subsection~\ref{subsec:The-dark-state-lambda}, linking the localization length $L_{\textrm{\textrm{loc}}}$ to the correlations functions of the potential. We discuss the possibility for approximation of the potential by properly placed Dirac-delta scatterers. 

In Section \ref{section:RandomComb} we analyze Anderson localization in potential defined in~\ref{subsec:The-dark-state-running-lambda}. We also discuss the role of correlations between potential peaks height on singularity in a dependence of the localization length on energy.

\begin{figure}
\includegraphics[width=8.6cm]{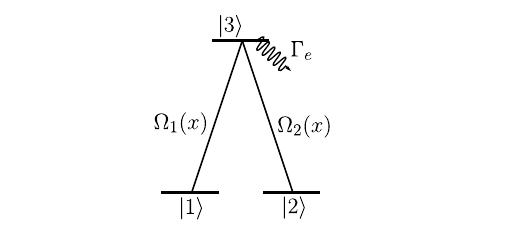}

\caption{The $\Lambda$  level configuration considered in this work. The states $|1\rangle,|2\rangle,$
are assumed to be the ground state sublevels while $|3\rangle$  is
the excited state with the spontaneous emission rate $\Gamma_{e}.$ The Rabi
frequencies $\Omega_{i}(x)$ may be due to laser standing waves or a
speckle field and are typically position dependent.}

\label{fig:5level}
\end{figure}

\section{The model}

\label{sec:model}

We consider a gas of ultracold atoms of mass $m$ confined to a 1D tube
along the $x$ axis by a tight transverse harmonic confinement in $y,z$
realised by the potential $V(x,y,z)=m\omega_{\perp}^{2}(y^{2}+z^{2})/2$.
The $\hbar\omega_{\perp}$ is assumed to be sufficiently large for
excited transverse modes to remain unpopulated.  We assume no confinement along the $x$ direction, but in real experiment one would use either harmonic confinement or sheet light implementing hard-wall boundary condition \cite{Gaunt2013}.

The atoms are driven by resonant laser light coupling three  atomic (sub)levels
in the $\Lambda$ configuration as shown in Fig.~\ref{fig:5level}.
We assume the gas is non-interacting. In \citep{Wang2018} this was realised using fermionic ${}^{171}$Yb, where $s$-wave contact interactions were suppressed. Whether the scheme can be successfully implemented with bosons is still an open question, due to possible detrimental effect of collisional losses.

The Hamiltonian of the model takes the form
\begin{eqnarray}
H & =- & \frac{\hbar^{2}}{2m}\frac{d^{2}}{dx^{2}}+H_a,\label{eq:ham3level}\\
H_a & = & \frac{\hbar}{2}
\left(\begin{array}{ccc}
0 & 0 & \Omega_{1}^*(x)\\
0 & 0 & \Omega_{2}^*(x)\\
\Omega_{1}(x) & \Omega_{2}(x) & -i\Gamma_{e}
\end{array}\nonumber\right).
%\label{eqn:ham3level}
\end{eqnarray}
The Rabi frequencies $\Omega_{i}$, $i=1,2$  describe laser driving of the corresponding transitions between 
 basis states  $|i\rangle$ and $|3\rangle$. The $\Gamma_{e}$ denotes the spontaneous emission decay rate of the upper state in the $\Lambda$ scheme. The fields $\Omega_{i}(x)$ can be due to a laser standing wave or
a speckle field as discussed later on.

The ``atomic'' part of the Hamiltonian, $H_a$ for each $x\in\mathbb{R}$
has a zero eigenvalue with associated ``dark state'' eigenvector:
\begin{eqnarray}
|D(x)\rangle\!\! &\! =\! &\!\! \frac{-\Omega_{2}(x)|1\rangle\!+\!\Omega_{1}(x)|2\rangle}{\sqrt{|\Omega_{1}(x)|^{2}\!+\!|\Omega_{2}(x)|^{2}}}\!=\!\cos\alpha_x|2\rangle\!-\!\sin\alpha_x|1\rangle,\label{eq:darkState3}
\end{eqnarray}
The remaining eigenvectors $|B_{j}(x)\rangle,j=1,2$ are called bright states
since they have a nonzero contribution from the excited state $|3\rangle$.
When $\Gamma_{e}\neq0$ the matrix $H_a$
is non-Hermitian and the set of right eigenvectors ${\cal B}=\{|D(x)\rangle,|B_{1}(x)\rangle,|B_{2}(x)\rangle\}$
has the associated ``ket'' states $\langle D(x)|,\langle B_{j}(x)|$ that complete the biorthonormal system.
The latter are always meant as  proper ``left'' eigenvectors, and in
general $\langle B_{i}(x)|\neq|B_{i}(x)\rangle^{\dagger}.$

The bright state energies are:
\begin{equation}
E_{j}(x)=\frac{\hbar}{4}\left(-i\Gamma+(-1)^j\sqrt{-\Gamma_{e}^{2}+4|\Omega_{1}(x)|^{2}+4|\Omega_{2}(x)|^{2}}\right).\label{eq:3evelEplus}
\end{equation}
The gap to bright
states is non-zero if both $\Omega_1(x), \Omega_2(x)$ do not vanish at some $x$. This can be ensured e.g. when  one of $\Omega_1(x)$ is position-independent and non-zero. 

When expressed in the position-dependent basis ${\cal B}$, defined above, the Hamiltonian  (\ref{eq:ham3level})
takes the form (see \citep{Wilczek1984,Ruseckas2005,dalibard2011colloquium}):
\begin{eqnarray}
H & = & \frac{1}{2m}(p-A)^{2}+\sum_{i=1}^2 E_{j}(x)|B_{j}\rangle\langle B_{j}|\label{eq:eq3channel}
\end{eqnarray}
with $A_{ij}=-i\hbar\langle{\cal B}_{i}|\partial_{x}|{\cal B}_{j}\rangle.$
One can always choose the local phases of basis vectors $|1\rangle,\ldots,|3\rangle$
such that $\Omega_{i}(x)$ are real. Then, after projection onto 
 $|D(x)\rangle$ state the Hamiltonian  (\ref{eq:ham3level})
reduces to  the form:
\begin{eqnarray}
H & = & -\frac{\hbar^{2}}{2m}\frac{d^{2}}{dx^{2}}+V_{D}(x),\label{eq:darkStateHamiltonian}
\end{eqnarray}
where 
\begin{eqnarray}
V_{D}(x) & =&-\frac{\hbar^{2}}{2m}  \langle D(x)|\partial_{xx}|D(x)\rangle,\label{eq:darkPotential}
\end{eqnarray}
is the dark state potential. Using Eq.~\eqref{eq:darkState3} one obtains:
\begin{equation}
V_D(x)=\frac{\hbar^2}{2m}\frac{(\Omega_1'(x)\Omega_2(x)-\Omega_1(x)\Omega_2'(x))^2}{(\Omega_1^2(x)+\Omega_2^2(x))^2}=\frac{\hbar^2}{2m} (\alpha_x')^2.
\label{eq:darkPotentialExplicite}
\end{equation}
Under the condition 
\begin{eqnarray}
|E_{j}(x)| & \gg & |A_{kl}(x)|,\quad j\in\{1,2\},\ k\neq l,\label{eq:darkAdiabatic}
\end{eqnarray}
valid for sufficiently large   $\Omega_i$  \cite{Lacki2016,Lacki2019}, 
the dark state is only very weakly depopulated.

\paragraph*{The Rabi frequencies.-}The Rabi frequencies in the Hamiltonian \eqref{eq:ham3level} considered in this work are due to a standing/running laser wave or a speckle field.  
In the former case they are of the form:
\begin{equation}
  \Omega_i(x)=\tilde{\Omega}_i\sin(k_ix+\phi)+\tilde{\Omega}_i^0.
  \label{eqn:OmegaY}
\end{equation}
The $k_i=2\pi/\lambda_i$ and $\lambda_i$ is the wavelength of the laser implementing $\Omega_i(x)$. The value of $k_i$ in Eq.~\eqref{eqn:OmegaY} may be also (smoothly) controlled if the lasers creating the standing wave propagate at a finite angle with respect to $\hat x$. The intensity of the lasers controls the amplitude $\tilde{\Omega}_i$. Implementation of the term $\tilde{\Omega}_i^0$ requires phase coherent projection of a running wave in the direction perpendicular to the $\hat{x}$ axis (see \cite{Yang2018}). 

The wave number $k_1$ defines the recoil energy:
\begin{equation}E_r=\frac{\hbar^2k_1^2}{2m}.
\label{eq:Er}
\end{equation}
in this work we always use the recoil energy defined with respect to $k_1$.
Thus $E_r$ carries no index "$i$".

The potential $V_{D}(x)$ is randomized by using random Rabi frequency
$\Omega_{i}(x)$.  That may be accomplished by driving the corresponding transition with a quasi-random electric
field in the form of the speckle field. It is created by propagating a laser beam from the direction perpendicular to the $x$ axis through a diffusive plate, and focusing it with the lens. The complex amplitude of the
electric field along the system, near the focal point of the lens, is then given by the formula \citep{Piraud2012}:
\begin{equation}
F(x)\sim \frac{e^{i\frac{2\pi f}{\lambda}}}{i\lambda f}e^{i\frac{\pi}{\lambda f}x^{2}}\int_{-R/2}^{R/2}d\rho\mu(\rho)w(\rho)e^{i\frac{\pi}{\lambda f}\rho^{2}}e^{-i\frac{2\pi}{\lambda f}x\rho}.\label{eq:speckle}
\end{equation}
The $\lambda$ is the laser wavelength, $f$ -- the focal distance
of the used lens and $R$ indicates the {radius} of the diffusive plate (we assume it to be identical to the radius of the lens).
Here we skip the index $i$. The $\mu(\cdot)$ are random complex phases imprinted by the
diffusive surface. They are assumed to be completely random phase
factors with a homogeneous probability density over a unit circle. The
above formula is valid in the paraxial approximation, namely $f\gg R$.
The ratio $R/f$ determines the degree to which the laser field is
focused. This ratio controls the effective length scale of $F(x)$ \eqref{eq:speckle}.
Specifically 
\begin{eqnarray*}
\sigma_{R} & = & \frac{\lambda f}{\pi R},
\end{eqnarray*}
is the correlation length for the speckle potential, a convenient length unit for the speckle field. In this work even if several speckle fields are used simultaneously in some laser configuration, it is assumed, for simplicity, that they have the same $\sigma_R$.
We then define
\begin{equation}
E_{\sigma_{R}}=\frac{\hbar^{2}}{2m\sigma_{R}^{2}},
\label{eqn:EsigmaR}
\end{equation}
as a characteristic speckle energy scale.  It is interesting to compare the above expression to the recoil energy for a laser with same wavelength. It is 
\begin{equation} 
E_r=(2f/R)^2E_{\sigma_R},
\end{equation}
which for the assumed in this work ratio $R/f=1/3$, leads to $E_r=36E_{\sigma_R}$. 
The field $F(x)$ generates a Rabi frequency ${\Omega}_i(x)$ which may be for convenience expressed as a product of its mean value $\tilde{\Omega}_i$ and the dimensionless function $S_i(x)$: 

\begin{equation}
\Omega_{i}(x)=\tilde{\Omega}_{i}S_i(x),
\label{eqn:Ei}
\end{equation}
where $\frac{1}{L}\int_{0}^{L}|S(x)|dx\to 1$ as $L\to\infty$. 
The $\Omega_i$ as above is non-zero, but it takes arbitrary small value with a finite probability. To overcome this problem (recall small $\Omega_i$ may be harmful to our $\Lambda$ scheme properties)  one can add a phase coherent laser field which leads to:
\begin{equation}
\Omega_{i}(x)=\tilde{\Omega}_{i}S_i(x)+\tilde{\Omega}_i^0,
\label{eqn:Ei2}
\end{equation}
where both $\tilde{\Omega}_i^0,\tilde{\Omega}_i$ are independently controlled by intensity of the respective laser field.
Again, without a loss of generality $S_i(x),\tilde{\Omega}_i^0,\tilde{\Omega}_i\in\mathbb{R}$.

\paragraph*{The Speckle potential.-} The speckle laser field can be used to create an optical speckle potential via the AC Start shift in the two level system. The speckle laser field with Rabi frequency $\Omega(x)$, detuned by $\Delta$ from the resonance creates the optical potential

\begin{equation}
V_{sp}(x)=\hbar\frac{\Omega^2(x)}{4\Delta}.
\label{eq:Vsp}
\end{equation}

\paragraph*{The Window function.-}The formula (\ref{eq:speckle}) takes into account the window function
$w(\cdot)$ which can be used to tune the statistical properties
of $F(x)$. We consider windows of the form 
\begin{equation}
w(\rho)=\Theta(|\rho|-R/2+W)-\Theta(|\rho|-R/2)\label{eq:window}
\end{equation}
which form a double-slit system \citep{Piraud2012}. In the simplest
case, $W=R/2$, $w(\rho)=1$ for $|\rho|\leq R/2.$ 

\section{The dark state potential }
\label{sec:dark-state-potential}

The features of the dark state potential $V_{D}(x)$ depend solely on those of the dark state  $|D(x)\rangle$, compare \eqref{eq:darkPotential}.
In contrast to  potentials created by  AC-Stark shifts,
tuning the laser intensity does not necessarily modify the amplitude of the potential.
Scaling of all $\Omega_i$ by a common factor leaves the dressed states and $V_D(x)$ unaffected due
to a functional form of $|D(x)\rangle$ [see \eqref{eq:darkState3}]. 

The amplitude and shape of $V_D(x)$
is rather controlled by relative magnitudes of the two Rabi frequencies $\Omega_1(x)$ and $\Omega_2(x)$, which prompts us to define the dimensionless parameter:
\begin{equation}
\epsilon_{12}=\frac{\tilde{\Omega}_1}{\tilde{\Omega}_2}=\epsilon_{21}^{-1},
\label{eqn:epsilon}
\end{equation}
controlling that aspect of the setup. Obviously when in a specific situation roles of $\Omega_1(x)$ and $\Omega_2(x)$ are interchangeable, then configurations for $\epsilon_{12}=\epsilon$ and $\epsilon_{12}=\epsilon^{-1}$ are equivalent.
From Eq.~\eqref{eq:darkPotential} one sees that potentials peaks in $V_D(x)$ occur where $|D(x)\rangle$ changes substantially over a short distance. This may occur e.g in those places where $\Omega_1(x),\Omega_2(x)$ go from  $\Omega_1(x)\ll\Omega_2(x)$ to  $\Omega_1(x)\gg\Omega_2(x)$ regime or vice versa.

To get more insight into the genesis and shape of $V_D(x)$, we first look in more detail at two important special cases. 

First, when $\Omega_2(x)$ is due to a speckle field and $\Omega_1(x)$ is constant or slowly varying on a scale much larger than the wavelength of the speckle.  The potential typically consists of double-peak structures that appear near minima of $\Omega_2(x)$. This is discussed below in \ref{subsec:The-dark-state-lambda} together with basic statistical properties of this potential.

Secondly, we consider the case when $\Omega_1(x)$ is due to a running wave, Eq.~\eqref{eqn:OmegaY}. Then $V_D(x)$ has sharp potential peaks near zeros of $\Omega_1$, where $\Omega_2(x)$
may be considered constant locally. To randomize the heights of $V_D(x)$ peaks, the $\Omega_2(x)$ may come from a speckle field or a running wave, Eq.~\eqref{eqn:Ei2} with a wavelength incommensurate with the $\Omega_1(x)$, creating a quasiperiodic pattern.

\subsection{The $V_D(x)$ near finite minima of $\Omega_2$ due to a speckle field }
\label{subsec:The-dark-state-lambda}

\begin{figure}
\includegraphics[width=8.6cm]{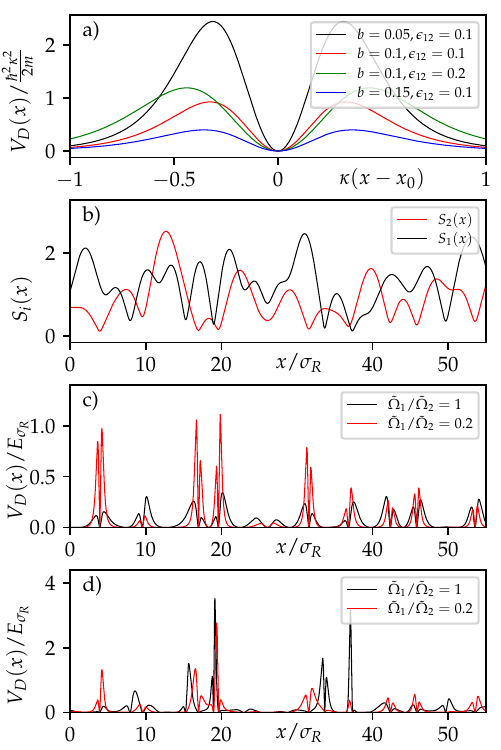}

\caption{Panel a): shape of double peak structure of $V_{D}(x)$ potential for the 3-level $\Lambda$
system near quadratic minimum of $\Omega_{2}(x)$ -- Eq.~(\ref{eq:omega1quadratic}). To reach a substantial peak height, one needs $b\ll\epsilon_{12}\ll1$. Panel b) shows two exemplary realizations of the speckle shape functions $S_i(x)$. Panels c) and d) show the dark state potential $V_D(x)$ when $\Omega_2(x)=\tilde{\Omega}_2 S_2(x).$ Panel c) is for a constant  $\Omega_1(x)=\tilde{\Omega}_1$ while in panel d) $\Omega_1(x)=\tilde{\Omega}_1S_1(x)$. The relative strengths of $\Omega_1(x)$ and $\Omega_2(x)$ are indicated in the legends.
}
\label{fig:lambdaDarkPotential}
\end{figure}

The $\Omega_{2}(x)$ coming from the speckle field does not feature
exact zeros, but rather it has local minima. Consider a minimum of $\Omega_2(x)$ at $x=x_0$. For $x\approx x_0$ 
we approximate $\Omega_2(x)$ as:
\begin{equation}
\Omega_{2}(x)\equiv\tilde \Omega_{2}\left[b+\frac{\kappa^2}{2}(x-x_{0})^{2}\right].\label{eq:omega1quadratic}
\end{equation}
The part in bracket is a quadratic expansion of the function $S_2(x)$ around a particular minimum. We do not include the value of $b$ in $\tilde{\Omega}_2$ as we assume that the $\tilde{\Omega}_2$ is defined by Eq.~\eqref{eqn:Ei2} for a given realization of $\Omega_2(x)$.

We consider $\Omega_1(x)=\tilde{\Omega}_1=\textrm{const.}$ and $|\Omega_{2}(x_0)|\ll\tilde{\Omega}_1$. Under these assumptions, the dark state potential $V_D(x)$ reveals, locally, a double peak structure
(see Fig.~\ref{fig:lambdaDarkPotential}a). Analytically, we have (see \cite{Yang2018}):
\begin{equation}
V_{D}(x)=\frac{\hbar^2\kappa^2}{2m}\frac{\epsilon_{12}^{2}\kappa^2(x-x_{0})^{2}}{\left[[b+\frac{\kappa^{2}}{2}(x-x_{0})^2]^{2}+\epsilon_{12}^{2}\right]^{2}},\label{eq:darkDoublePeak}
\end{equation}
where $\epsilon_{12}$ is given by Eq.~\eqref{eqn:epsilon}. The value of this parameter depends only on amplitudes of the Rabi frequencies, and is the same for different minima of a single realization of $\tilde{\Omega}_1(x)$. 

For arbitrary $\epsilon_{12},b$ the width of this structure is \begin{equation}
\Delta x(b,\epsilon_{12},\kappa)=2\kappa^{-1}\sqrt{\frac{2}{3}}\sqrt{\sqrt{4b^{2}+3\epsilon_{12}^{2}}-b}
\end{equation}
and its height is 
\begin{equation}
V_{\textrm{max}}(b,\epsilon_{12},\kappa)=\frac{\hbar^{2}\kappa^{2}}{2m}\frac{27\epsilon_{12}^{2}\left(\sqrt{4b^{2}+3\epsilon_{12}^{2}}-b\right)}{8\left(b\left(\sqrt{4b^{2}+3\epsilon_{12}^{2}}+2b\right)+3\epsilon_{12}^{2}\right)^{2}}.\end{equation}

For $\epsilon_{12} \gg b$ the width:
\begin{equation}
\Delta x(b,\epsilon_{12},\kappa) \to \frac{\sqrt{8\epsilon_{12}}}{\sqrt[4]{3}}\kappa^{-1},
\end{equation}
and the height : 
\begin{equation}
V_{\textrm{max}}(b,\epsilon_{12},\kappa) \to \frac{3\sqrt{3}}{8\epsilon_{12}}\frac{\hbar^{2}\kappa^{2}}{2m}.
\label{eq:Vdp}
\end{equation}
If additionally  $\epsilon_{12}\to0$ the two potential peaks converge to $\frac{\pi}{2\sqrt{\epsilon_{12}}}\delta(x-x_{0})$. 

For $b\gg\epsilon_{12}$ the width is 
\begin{equation}
\Delta x(b,\epsilon_{12},\kappa)\to 2k^{-1}\sqrt{\frac{2}{3}b},\end{equation}
and  the height of both peaks is 
\begin{equation}
V_{\textrm{max}}(b,\epsilon_{12},\kappa) \to \frac{27\epsilon_{12}^{2}}{128b^{3}}\frac{\hbar^{2}\kappa^{2}}{2m}.
\label{eq:Vdp2}
\end{equation}

The Figure~\ref{fig:lambdaDarkPotential}c) shows the exemplary dark state potentials obtained for $\Omega_2(x)$ equal to the speckle shown in Fig.~\ref{fig:lambdaDarkPotential}b) with a black line, while $\Omega_1$ remains position independent. The plot shows two cases $\epsilon_{12}=1$ and $\epsilon_{12}=0.2$ with relative peak heights following~\eqref{eq:Vdp} and~\eqref{eq:Vdp2}. For fixed $\epsilon_{12}$ we may ascribe the value of parameter $b_i$ from Eq.~\eqref{eq:omega1quadratic} to each of the minima of $\Omega_2(x)$ at $x_i$, indexed by $i$. If the value of $\epsilon_{12}$ is lowered, potential peaks for which $\epsilon_{12}\gg b_i$ are made higher and narrower, but those that already passed to the opposite $\epsilon_{12}\ll b_i$ regime have their height further reduced (see Eq.~\eqref{eq:Vdp2}). Decrease of $\epsilon_{12}$ results in fewer sharp peaks in $V_D(x)$ but height of some of those peaks can increase.

Similar observations may be made in the case when $\Omega_1(x)$ is not constant but is due to a speckle field itself. Figure~\ref{fig:lambdaDarkPotential}d) shows the corresponding exemplary potential for same $\Omega_2(x)$ as in Fig.~\ref{fig:lambdaDarkPotential}c) and $\Omega_1(x)$ given by the red curve in Fig.~\ref{fig:lambdaDarkPotential}b). Most of the potential peaks occur where one of $\Omega_1(x)$, $\Omega_2(x)$ has a minimum and the other may be considered approximately constant. Similarly one can use Eq.~\eqref{eq:omega1quadratic} applied to $\Omega_1(x)$ or $\Omega_2(x)$ and ascribe $b_i$'s to each  minimum. 

Since now $\Omega_1(x)$ is position dependent, in order  to characterize individual peaks near minima of $\Omega_2(x)$ via~\eqref{eq:darkDoublePeak} we have to substitute $\epsilon_{12}\to\epsilon_{12,i}$ where 
\begin{equation}
\epsilon_{12,i}=\frac{\tilde{\Omega}_1S_1(x_i)}{\tilde{\Omega}_2},
\end{equation}
with $\epsilon_{12,i}$  specific for each minimum. In case of the minima of $\Omega_1$, we consider $\epsilon_{21,i}$ defined as above with swapped $\Omega_1$ and $\Omega_2$.   

Let us consider reducing the amplitude $\tilde{\Omega}_1$. As $\epsilon_{12,i}\sim\tilde{\Omega}_1$, the discussion of regimes $\epsilon_{12,i}\ll b_i$ vs $\epsilon_{12,i}\gg b_i$ carried out for constant $\Omega_1(x)$ still applies. The potential peaks near the minima of $\Omega_1(x)$ are characterized by $\epsilon_{21,i} \sim \tilde{\Omega}_1^{-1}$. Thus for smaller and smaller $\tilde{\Omega}_1$ height of the latter family of peaks is reduced as well.   

\begin{figure}[th]
\includegraphics[width=8.6cm]{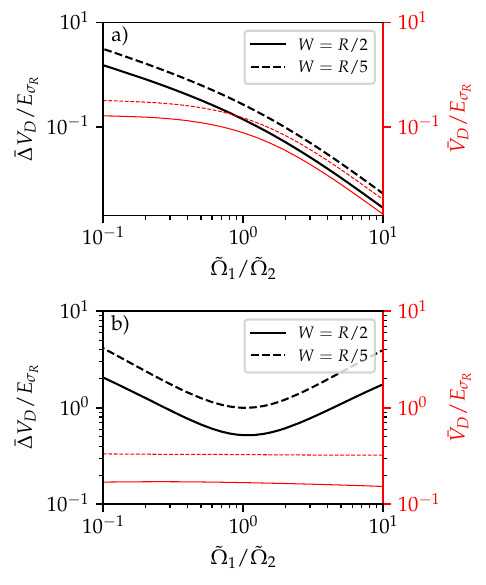}

\caption{ Panel (a) shows the mean height $\bar{V}_D$ (red lines) and standard
deviation $\bar{\Delta}V_D$ (black lines) for potential $V_{D}(x)$ for 
$W=R/2$ (solid) and  $W=R/5$ (dashed lines). In both panels $\Omega_2(x)=\tilde{\Omega}_2S_2(x)$. In panel a) $\Omega_1(x)=\tilde{\Omega}_1$ and in b) $\Omega_1(x)=\tilde{\Omega}_1 S_1(x)$ is due to a speckle field.}

\label{fig:potentialHeightdistro}
\end{figure}

Let us now see how the above observations manifest in statistical properties of the potential $V_D(x)$. Figure~\ref{fig:potentialHeightdistro}  presents $\bar V_D$ -- the mean, and $\bar{\Delta} V_D$ -- the standard deviation of $V_D(x)$ as a function of $\tilde{\Omega}_1/\tilde{\Omega}_2$ for the case of constant $\Omega_1(x)=\tilde{\Omega}_1$ (panel a) and for the case when $\Omega_1(x)=\tilde{\Omega}_1 S_1(x)$ (panel b). In the latter there is an obvious symmetry $(\tilde{\Omega}_1,\tilde{\Omega}_2) \to (\tilde{\Omega}_2,\tilde{\Omega}_1)$.

In both cases for $\tilde{\Omega}_1/\tilde{\Omega}_2 < 1$, as this ratio decreases, the standard deviation of $V_D(x)$ grows, and mean $\bar{V}_D$ converges to a constant. This is consistent with increasingly more sparse minima satisfying $b_{i}\ll \epsilon_{12}$ (or $b_i\ll \epsilon_{12,i}$ for panel b) ).

For large values of $\tilde{\Omega}_1/\tilde{\Omega}_2$, in case of constant  $\Omega_1(x)=\tilde{\Omega}_1 \gg \Omega_2(x)$, we approximately have:
\begin{equation}
|D(x)\rangle \approx -\frac{\Omega_2(x)}{\tilde{\Omega}_1}|1\rangle + |2\rangle, \end{equation}
 and: 
\begin{equation}
V_D(x) \approx \frac{\Omega_2'(x)^2}{\tilde{\Omega}_1^2}=(S_2'(x))^2 \epsilon_{12}^{-2}.
\label{eq:VdNotlattice}
\end{equation}
This means that both the mean  height $\bar{V}_D$ and the standard deviation ${\bar{\Delta}}V_D$ decrease  to 0, for incraesing $\tilde{\Omega}_1/\tilde{\Omega}_2$.  Their ratio $\bar{\Delta}V_D/\bar{V}_D \to 1.32\pm 0.02$  as seen in Fig.~\ref{fig:potentialHeightdistro}a). This limit is larger than the $\bar{\Delta}V_{sp}/\bar{V}_{sp} \to 1$ for the far-detuned AC-Stark optical potential $V_{sp}(x)$ created by laser speckle, in a standard optical lattice setting where $V_{sp}(x)\sim \Omega_1^2(x)/(4\delta)$ (with $\delta$ the detuning from the resonance).

In the situation when both $\Omega_1(x)$ and $\Omega_2(x)$ are due to speckle fields, the standard deviation ${\bar{\Delta}}V_D$ decreases towards a minimum at exactly $\tilde{\Omega}_1=\tilde{\Omega}_2$. For both $\tilde{\Omega}_1/\tilde{\Omega}_2 \to 0$ and $\tilde{\Omega}_1/\tilde{\Omega}_2\to \infty$ the behaviour of $\bar{\Delta}V_D$ is similar to the case of constant $\Omega_1(x)$. The marked difference is that $\bar{V}_D$ is $\tilde{\Omega}_1/\tilde{\Omega}_2$-independent. Qualitatively  speaking, this is because change of $\tilde{\Omega}_1/\tilde{\Omega}_2$ has the opposite effect on potential peaks near minima of $\Omega_2(x)$ and $\Omega_1(x)$ when it comes to their height and width.

In Fig.~\ref{fig:potentialHeightdistro} we mark with the 
  dashed lines results for  two cases discussed above when the obstacle is put onto the diffusive plate. We chose to illustrate this  by setting the parameter $W=R/5$, in Eq.~\eqref{eq:speckle} (note that the case $W=R/2$ corresponds to no obstacle). The obstacle suppresses low frequencies from the Fourier expansion of the $F(x)$ and the resulting potential $V_D(x)$ has higher mean and variance.

\subsection{The dark state potential near zeros of $\Omega_i$'s }
\label{subsec:The-dark-state-running-lambda}

Let us now consider the situation when $\Omega_{1}(x)$ posses a zero
over the real axis at $x=x_i$, as  in, e.g., the case of $\Omega_1(x)$ being due to a standing wave, Eq.~\eqref{eqn:OmegaY}. We assume $\Omega_2(x)$ to be locally constant $\Omega_2(x)\approx \tilde{\Omega}_{2,i}$ near $x_i$. This creates the setting similar to the dark state lattice proposal \cite{Lacki2016}. We then linearize

\begin{figure}
\includegraphics[width=8.6cm]{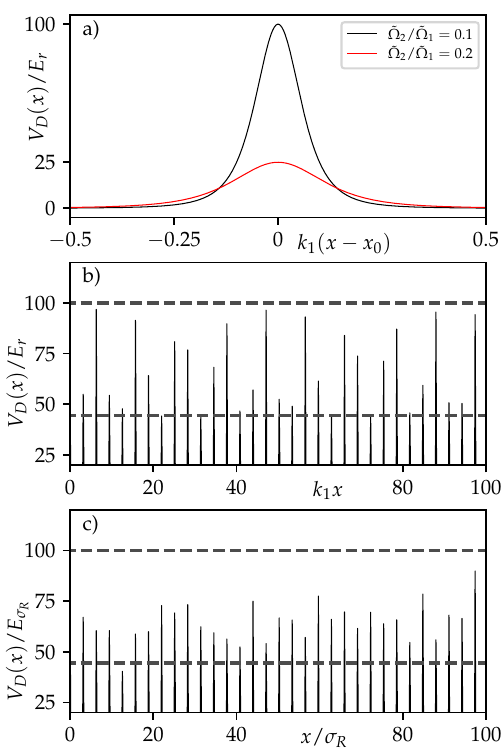}

\caption{ Panel a): shape of peak structure in $V_{D}(x)$ potential for the 3-level $\Lambda$
system near a zero of $\Omega_{1}(x)$ at $x=x_0$- Eq.~(\ref{eq:Vdlinear}) for different values of $\tilde{\Omega}_2/\tilde{\Omega}_1$. Panel b) shows $V_D(x)$ when both $\Omega_1(x), \Omega_2(x)$ are due to a standing wave. The $\Omega_1(x)=\tilde{\Omega}_1\sin(k_1 x)$ and $\Omega_2(x)=\tilde{\Omega}_2\sin(k_2 x)+\tilde{\Omega}_2^0$ such that $\epsilon_+=0.15,\epsilon_-=0.1$. The ratio $k_1/k_2$ takes the examplary value $k_1/k_2=(1+\sqrt{5})/2$. Panel c) dark state potential $V_D(x)$ for the $\Lambda$ system with $\Omega_1(x)=\tilde{\Omega}_1\sin(k_1 x),$ $\Omega_2(x)=\tilde{\Omega}_2S_2(x)+\tilde{\Omega}_2^0$ such that $\epsilon_+=0.15,\epsilon_-=0.1$. The $k_1$ is set such that $k_1\sigma_R = 1$, and $R/f=1/3$. 
}
\label{fig:lambdaDarkPotentialSin}
\end{figure}
\begin{equation}
\Omega_{1}(x)\approx\tilde{\Omega}_{1}k_1(x-x_{0})\label{eq:omega1linear},
\end{equation}
which gives  $V_{D}(x)$ of the form:
\begin{equation}
V_{D}(x)\approx\frac{\epsilon_{21,i}^{2}E_{r}}{[k_1^{2}(x-x_{0})^{2}+\epsilon_{21,i}^2]^{2}}
\label{eq:Vdlinear}
\end{equation}
It describes a peak of width $\sim\epsilon_{21,i}\lambda_1$ and
height $\epsilon_{21,i}^{-2}E_{r}$ with $\epsilon_{21,i}=\tilde{\Omega}_{2,i}/\tilde{\Omega}_1$
(see Fig.~\ref{fig:lambdaDarkPotentialSin}a). In the limit $\epsilon_{21,i}\to 0$ each of the potential peaks converges to $\frac{\pi}{2\epsilon_{21,i}}\delta(x-x_{0})$. If $\Omega_2(x)$ were truly constant, the subsequent peaks would create a lattice of narrow peaks of identical shape and height, just as in \cite{Lacki2016}.  To randomize them, we use pseudorandom  $\Omega_{2}(x)$. We discuss two possibilities. 

One option is to choose $\Omega_2(x)$ as in Eq.~\eqref{eqn:OmegaY} with $k_1/k_2\neq\mathbb{Q}$. In that case for different $x_i$ such that $\Omega_1(x_i)=0$ we have $\epsilon_{21,i}$ that vary between $\epsilon_-=\max(0,[-\tilde{\Omega}_2+\tilde{\Omega}_2^0]/\tilde{\Omega}_1)$ and $\epsilon_+=[\tilde{\Omega}_2+\tilde{\Omega}_2^0]/\tilde{\Omega}_1.$ This translates into pseudo-random height and width of subsequent peaks of $V_D(x)$ determined by subsequent $\epsilon_{21,i}$'s. The resulting potential consisting of equidistant pseudorandom peaks is shown in Fig.~\ref{fig:lambdaDarkPotentialSin}b) for $\epsilon_+=0.15$ and $\epsilon_-=0.1$. The expressions for $\epsilon_-$, $\epsilon_+$ show that one can control the amplitude of the disorder simply by changing $\tilde{\Omega}_2$, $\tilde{\Omega}_2^0$ and $\tilde{\Omega}_1$

One should note that, in general, there are additional potential peaks near minima of $\Omega_2(x)$ at points designed $x'_i$. Such peaks are described by Eq.~\eqref{eq:Vdlinear} or Eq.~\eqref{eq:darkDoublePeak} with values of $\epsilon_{12,i}=\Omega_1(x'_i)/\Omega_2(x'_i)$ for $x_i'$ far from any $x_j$ we have $\epsilon_{12,i}\gg $ and for $x'_i$ equal to some $x_j$ the potential peak is mainly due to zero of $\Omega_1(x)$. These peaks are automatically included in the numerical treatment of the model that takes exact value of $V_D(x)$.

One can make similar construction with $\Omega_2(x)$ due to a speckle field, Eq.~\eqref{eqn:Ei2}. In contrast to the sine function case, the $S_2$  in Eq.~\eqref{eqn:Ei2} is strictly limited only from below (by zero). The probability for taking the value above 2 is nevertheless exponentially suppressed. This means that for most $\epsilon_i$ characterizing individual peaks, we have $\epsilon_i\in[\epsilon_-,\epsilon_+]$ where  $\epsilon_-=\max(0,[-2\tilde{\Omega}_2+\tilde{\Omega}_2^0]/\tilde{\Omega}_1)$ and $\epsilon_+=[2\tilde{\Omega}_2+\tilde{\Omega}_2^0]/\tilde{\Omega}_1.$  The resulting potential $V_D(x)$ is shown in Fig.~\eqref{fig:lambdaDarkPotentialSin}c) for $\epsilon_+=0.15$ and $\epsilon_-=0.1$. In broad terms it is similar to the previously considered $\Omega_2(x)$ as in Eq.~\eqref{eqn:OmegaY}, but differs in statistical properties of peak heights. This is discussed further in Section \ref{section:RandomComb} where we calculate tight-binding parameters for movement in this kind of random potential.

\section{Localization}
\label{sec:localization}

In the case when $\Omega_1(x)=\textrm{const.}$ and $\Omega_2(x)=S_2(x)\tilde{\Omega}_2$ the $V_D(x)$ consists of relatively narrow random double peaks.    In this settings it is natural to consider Anderson localization which has been traditionally studied in the optical potential created by a speckle field via AC-Stark effect. To that end we first discuss the two-point correlation function of the $V_D(x)$ potential in such a case.

\subsection{Correlation functions}

Let us consider two point 
correlation function
\[
C_{2}(\delta x)=\overline{V(x)V(x+\delta x)}.
\]
For random potentials it 
is directly related to the so-called Anderson 
localization length, $L_{\textrm{loc}}$
of the eigenstates \citep{Lugan2009,Piraud2012}. 
Generically in one-dimensional systems with a random potential $V(x)$,
\begin{eqnarray}
H & = & -\frac{\hbar^{2}}{2m}\frac{d^{2}}{dx^{2}}+V(x),\label{eq:mainH}
\end{eqnarray}
one expects \citep{Anderson1958,Lugan2009} that eigenstates $\psi_{i}(x)$ are exponentially
localized:
\begin{eqnarray}
|\psi_{i}(x)| & \sim & \exp[-|x|/L_{\textrm{loc}}(E)].\label{eq:psiexp}
\end{eqnarray}
The $E$ in $L_{\textrm{loc}}(E)$ is the energy of $\psi_i$. Often $L_{\textrm{loc}}(E)$ quickly grows with $E$.

The localization length may be related to the correlation function via a series expansion with respect to increasing powers of $(\bar{\Delta}V/\sqrt{E_{\sigma_R}(E-\bar{V})})^{1/2}$, where $\bar V,\bar{\Delta}V$ are the mean and standard deviation of $V(x)$. Specifically:
\begin{equation}
L_{\textrm{loc}}^{-1}(E-\bar{V})=\sum_{n\geq2}\gamma^{(n)}(E-\bar{V}).\label{eq:locCorrel}
\end{equation}
The lowest term $\gamma^{(2)}$ is given by
\begin{equation}
\gamma^{(2)}(E-\bar{V})=\frac{m}{4\hbar^2(E-\bar{V})} \tilde{C}_{2}\left[2\sqrt{\frac{2m(E-\bar{V})}{\hbar^2}}\right].\label{eq:CorrLoc}
\end{equation}
Here, $\tilde{C}_2$ is a Fourier transform of $C_2$.
Higher order terms contain multi-point correlation functions, beyond
two-point $C_{2}$. The expansion holds for a small $\bar{\Delta}V$. The other cases can be handled by numerical determination of $L_{\textrm{\textrm{loc}}}^{-1}$.

\begin{figure}
\includegraphics[width=8.6cm]{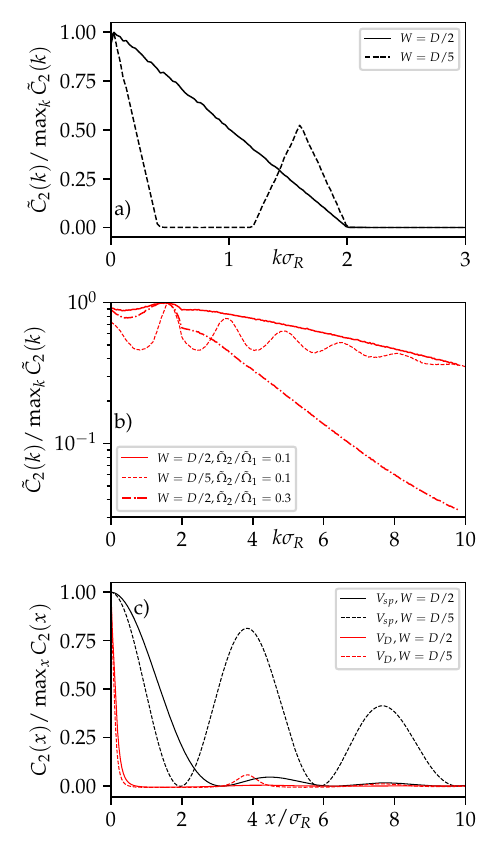}

\caption{Panel (a) shows $\tilde{C}_{2}$ for the speckle potential $V_{sp}$ for $W=D/2$
(solid line), $W=D/5$ (dashed line). Panel (b) same as above for
the dark state $V_{D}(x)$ potential with $\Omega_{2}(x)=\tilde{\Omega}_2S_2(x)$
-- a speckle field) for $\tilde{\Omega}_2/\tilde{\Omega}_1=0.1$ and $W=D/2$ (solid line), $W=D/5$ (dashed line). The dashed-dotted line shows data for $\tilde{\Omega}_2/\tilde{\Omega}_1=0.3$ and $W=D/2$.
Panel (c) shows the spatial correlation function $C_{2}(x)$  for
all of the above potentials with matching colors. In all of the above
the normalization of the plot has been chosen so that the maximal
value of each line is 1.}

\label{fig:corr}
\end{figure}

For the speckle optical potential, $V_{sp}(x)$, with a constant window function $W=R/2$,
the correlation function~is: 

\begin{equation}
C_{2}(\delta x)=\overline{V_{sp}}^{2}\left\{ 1+\left[\frac{\sin(x/\sigma_{R})}{x/\sigma_{R}}\right]^{2}\right\} ,
\end{equation}
and its Fourier transform (see Fig.~\ref{fig:corr}a):

\begin{equation}
\tilde{C}_{2}(k)=\overline{V_{sp}}^{2}\left\{ 2\pi\frac{\delta(k)}{\sigma_{R}}+\pi\max\left(0,1-\frac{|k|\sigma_{R}}{2}\right)\right\} .
\end{equation}

It is important to note that for $|k|\geq k_0=\frac{2}{\sigma_R}$  $\tilde{C}_2(k)$ vanishes \citep{Gurevich2009,Lugan2009}. This implies significantly longer localization
lengths for energies $E$ above $\bar{V}+E_{0},\ E_{0}=\frac{\hbar^{2}k_0^2}{2m}.$ This is because the value of $L_{\textrm{loc}}$ is solely due to higher order terms in the expansion (\ref{eq:locCorrel}).
 
The insertion of the obstacle in the optical system, that amounts to  $W\neq R/2$  in (\ref{eq:window}), has a profound impact on the correlation function $\tilde{C}_2(k)$. For $W\leq D/3$  the $\tilde{C}_{2}(k)$ vanishes not only for $|k|\geq k_0$ but also  for some intermediate values of $|k|$ within the interval $[0,k_0]$ as well. This is illustrated for $W=D/5$ in Fig.~\ref{fig:corr}a).

Consider now
the dark state potentials $V_D(x)$ as in the preceding section, for the case when $\Omega_1(x)=\tilde{\Omega}_1$ and $\Omega_2(x)=\tilde{\Omega}_2S_2(x).$  For such a configuration the  correlation functions $C_{2}$ and $\tilde{C}_{2}$ are shown the Fig.~\ref{fig:corr}b) and Fig.~\ref{fig:corr}c). Contrary to the $ V_{sp} $  potential case, here  $\tilde{C}_{2}(k)$ is non-zero for 
large values of momenta, $k$. This corresponds, in the position space, to the shape of $C_{2}$ shown in Fig.~\ref{fig:corr}c) where the dark-state
$C_{2}(x)$ features a narrow peak. 
These statements hold for both $W=R/2$ and $W=R/5$. In the latter case, when the obstacle is put in front of the diffusive plate, the strong modulation of $\tilde{C}_2(x)$ occurs.

Let us track the reason why high Fourier components $\tilde{C}_2$ behave differently for $V_{sp}$ and $V_D(x)$. The speckle potential $V_{sp}(x)$ is proportional to the square of the Rabi frequency $\Omega(x)$, as in Eq.~\eqref{eq:Vsp}. Taking the square at most doubles the extent of $k$ that index non-zero Fourier components of $V_{sp}$. This allows $\tilde{C}_2(k)=0,\ k\geq k_0$. 
In the case of the dark state potential, the highly nonlinear dependence of  $V_D(x)$ on $\Omega_i(x)$'s in Eq.~\eqref{eq:speckle} produces arbitrarily large Fourier components in $V_D(x)$ and there is no reason for $\tilde{C}_2(k)$ to vanish for large $k$. This is a manifestation of the origin of the dark state potential coming from position-dependent dark state in contrast to the conventional AC-Stark shift.  

Another feature worth pointing out is that by changing the ratio $\tilde{\Omega}_2/\tilde{\Omega}_1$ one controls the shape of the potential as proven by manifestly different $\tilde{C}_2$ for $\tilde{\Omega}_2/\tilde{\Omega}_1$ set to two exemplary values of 0.1 and 0.3. In case of the speckle potential change of $\Omega(x)$ changes the constant factor in $\tilde{C}_2(x)$, but keeps the overall shape of $\tilde{C}_2$ from Fig.~\ref{fig:corr}a).

\subsection{Anderson localization in a dark state potential}
\label{subsec:localization}
                      
To quantitatively analyze the physical implications of a particular form of $\tilde{C}_{2}$, we simulate the Anderson localization of a particle moving
in $V_{D},V_{sp}.$ Specifically, we look for eigenstates of Hamiltonian (\ref{eq:mainH}) at energy $E$ such that $\hbar^{2}k^{2}=2m(E-\bar{V})$. The resulting Schr\"odinger equation is solved over an interval $x\in[0,L]$ with the condition 
$\psi(x)\to e^{-ikx}, x\to 0+$. This is the outgoing amplitude of a particle that has entered the sample at $x=L$.  Near $x=L$ the wavefunction has the incoming and reflected components $\psi(x\to L)=Ae^{-ikx}+Be^{ikx}$ proportional to $A$ and $B$ respectively. The values of $A$, $B$ are determined numerically.

\begin{figure}
\includegraphics[width=8.6cm]{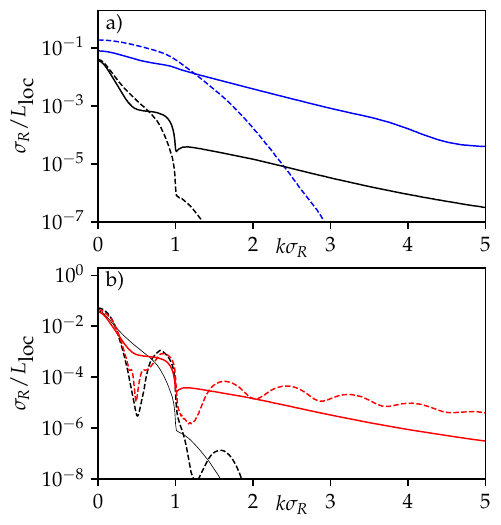}

\caption{ Localization lengths $L_{\textrm{loc}}$ for various considered potentials.
Panel a) compares the localization length in a speckle potential (dashed lines)
and $V_{D}$ potential for the $\Lambda$ system with $\Omega_1(x)=\tilde{\Omega}_1,\Omega_2(x)=\tilde{\Omega}_2S_2(x)$. For the speckle field we show two cases $\bar{V}=\bar{\Delta}V={0.04 E_{\sigma_{R}}}, 0.5 E_{\sigma_{R}}$ respectively with black dashed and blue dashed lines. Dark state potential $V_D(x)$ for matching $\bar{\Delta} V$ is shown with same colors and solid line. Respectively $\tilde{\Omega}_{1}/\tilde{\Omega}_{2}=\epsilon_{12}=2.357$ (black) and $\epsilon_{12}=0.333$ (blue).
Panel b) shows the effect of putting the obstacle in optical paths. The solid lines show $\sigma_R/L_{\textrm{loc}}$ for $W=R/2$ (lines repeated from a) for easing the comparison) and $W=R/5$ (dashed lines). The $\epsilon_{12}=2.357$ and $3.398$ ensure $\bar{\Delta} V=0.04E_{\sigma_R}$ for the $V_D(x)$ potential for $W=R/2$ and $W=R/5$ respectively.
}
\label{fig:loclength}
\end{figure}

We define the localization length $L_{\textrm{loc}}$ by the condition
\begin{eqnarray}
\langle\log|A|\rangle & \to & L/L_{\textrm{loc}}, \quad L\to \infty,
\label{eqn:locLength}
\end{eqnarray}
where $\langle\cdot\rangle$ denotes averaging over disorder realizations.

Figure~\ref{fig:loclength} shows $L_{\textrm{loc}}$  for large $L=5\times 10^{4}\sigma_{R},$ and $10^{4}$ disorder realizations. Let us focus on the black dashed curve corresponding
to $\sigma_{R}/L_{\textrm{loc}}$ for $V_{sp}$ with shallow $\bar{V}_{sp}=\bar{\Delta}V_{sp}=0.04 E_{\sigma_R}.$ Its dependence on $k$ shows a kink at $k=k_0$ such that $k_0\sigma_R=1$. By Eq.~\eqref{eq:CorrLoc} this corresponds to a  transition from $\tilde{C}_2 \neq 0$ for $k\leq 2k_0$ to $\tilde{C}_2=0$ for $k\geq 2k_0$. The kink is followed by a sudden increase of $L_\textrm{loc}$ as first observed in \cite{Lugan2009}.

We now show $\sigma_{R}/L_{\textrm{loc}}$ computed numerically
for the dark state potential $V_{D}.$ We focus on the case where $\Omega_1(x)=\tilde{\Omega}_1$ and $\Omega_2(x)=\tilde{\Omega}_2S_2(x)$ and present it in the same Fig.~\ref{fig:loclength}a). We chose the value of $\tilde{\Omega}_1/\tilde{\Omega}_2\approx 2.357$ to ensure that $\bar{\Delta} V_D=\bar{\Delta} V_{sp}=0.04E_{\sigma_R}$. One sees that for low momenta, smaller than the threshold value set by $k_0$, the localization length is similar to that for the speckle potential. At $k_0$ both potentials feature a kink. For such a small potential variance, the main contribution to the correlation length comes from $\gamma^{(2)}(k_0)$. It is proportional to $\tilde{C}_2(2k_0)$. For $k>k_0$  $\tilde{C}_2(k)=0$ for  the the speckle potential while for the dark state potential  a notable kink in $\tilde{C}_2$ at $k_0$ remains.

The main difference comes for $k\sigma_R \geq k_0 \sigma_R$ where the localization is  strongly suppressed in the speckle potential but not in the dark state potential $V_D$, again easily explained by properties of $\tilde{C}_2$. Thus the non-linear dependence of the potential $V_D$ on $\Omega$'s translated directly to an observable much stronger localization for large particle energies. 

For a sufficiently large amplitude of the disordered potential, the $\gamma^{(2)}$ term is no longer a dominant contribution to the inverse localization length. This is evident in Fig.~\ref{fig:corr}a) where $\bar{\Delta}V_D=\bar{\Delta}V_{sp}=0.5E_{\sigma_R}$. The kink at $k_0$ no longer can be observed in the dependence of $\sigma_R/L_{\textrm{loc}}$ on $k\sigma_R$ for both $V_{sp}$ and $V_{D}$, and the localization length is strongly decreased. Still for large momenta the localization is much stronger in the non-linear dark state potential.

When a non-trivial window function is used, the correlation function
$\tilde{C}_{2}$ for the $V_{D}$ potential, for increasing $k\sigma_{R}$
shows oscillations as in Fig.~\ref{fig:corr}b). These oscillations find
their way to the dependence of $\sigma_{R}/L_{\textrm{loc}}$
on the free momentum of the wave-function {[}see Fig.~(\ref{fig:loclength})b){]}.

\subsection{Dirac Delta approximation}

\label{subsec:Dirac-Delta-approximation}

\begin{figure}
\includegraphics[width=8.6cm]{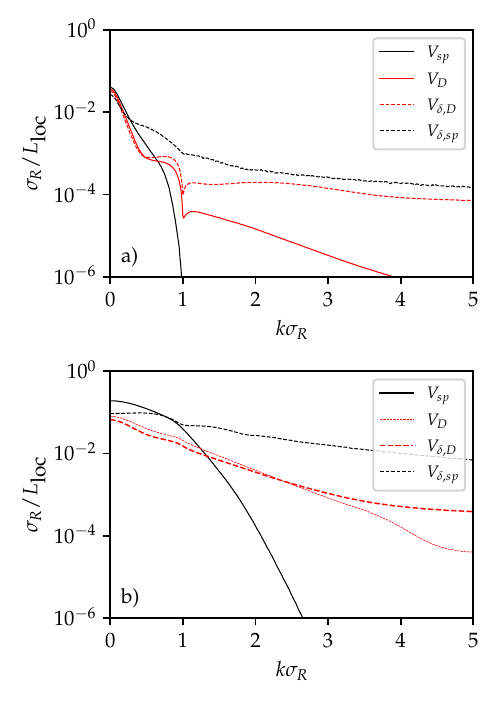}

\caption{Inverse localization length $\sigma_R/L_{\textrm{loc}}$ for potentials $V_{sp}$, $V_D$ and their Dirac delta approximations $V_{\delta,sp}$, $V_{\delta,D}$ (see legend) for [panel a)] shallow disorder: $\bar{\Delta}V_D=\bar{\Delta} V_{sp}=0.04 E_{\sigma_R}$ and [panel b)] $\bar{\Delta}V_D=\bar{\Delta} V_{sp}=0.5 E_{\sigma_R}$. In all cases the peak approximation parameter $\delta=0.25$.}

\label{fig:loclength2}
\end{figure}
 In this Section we determine if localization in the  dark state potential $V_D$ may be approximately described using a potential consisting of series of Dirac-delta peaks (Kronig-Penney model), $V_{\delta,D}(x)$:
 \begin{equation}
H=\frac{p^{2}}{2m}+\underbrace{\sum_{n}V_{n}\delta(x-x_{n})}_{V_{{\delta,D}}(x)}.\label{eq:kp}
\end{equation}
{Specifically, we compare the Anderson localization length for both $V_D$ and $V_{\delta,D}$.  }

To choose $V_{n}$ and $x_{n}$ for a particular potential
realization of $V_D(x)$ and obtain the approximate $V_{\delta,D}(x)$, we define a sequence of intervals $I_{n}=(a_{n},b_{n})\subset\mathbb{R}$
such that:
\begin{lyxlist}{00.00.0000}
\item [{(A)}] $V(x)$ has at least one local maximum in $I_{n},$
\item [{(B)}] $V(a_{n}),V(b_{n})\leq\delta\max_{x\in I_{n}}V(x)$ for $\delta$
being a small positive real number,
\item [{(C)}] no sub-interval contained in $I_{n}$ satisfies the above.
\end{lyxlist}
Intuitively, we want each intervals to contain a large portion of a single potential peak. The small value of $\delta$ ensures that the $V(x)$ is small outside of each interval $I_n$ with respect to the maximum value. On the other hand $\delta$ should not be chosen too small as it would lead to too large $I_{n}$ encompassing more than one peak. We opt to choose $\delta=0.25$.

The above definition does not automatically imply that different intervals
are disjoint. To ensure that, we actually find $I_{n}$ in the following
way:
\begin{enumerate}
\item For numerics we consider a particular realization over a finite interval $x\in[0,L]$.
\item We store all local maxima of $V(x),\ x\in[0,L]$ in the decreasing order with
respect to their value,
\item We find the interval $I_{1}$ encompassing largest maximum that satisfies
(A)-(C),
\item After first $n$ of intervals $I_{n}$ are determined, the (A)-(C)
define a candidate for the next interval $I_{n+1}'$. The set
$I_{n+1}:=I_{n+1}'\setminus\bigcup_{i=1}^{n}I_{i}$ is an interval. If it is empty then it is not added to the ${I_n}$ sequence. 
\end{enumerate}
 Each $I_{n}$ allows us to define an effective peak height
\begin{equation}V_{n}=\int_{I_{n}}V(x)\textrm{d}x,
\label{eqn:effHeight}
\end{equation}
and position
\begin{equation} x_{n}=\frac{1}{V_n}\int_{I_{n}}xV(x)\textrm{d}x.
\label{eqn:effPos}
\end{equation}

Let us note that it is possible that two very close maxima, for which $V(x)$ does not fall below the threshold defined by the $\delta$ will be approximated by a single Dirac Delta.

\paragraph*{Localization length calculation}

We have performed the transfer-matrix calculation of $\sigma_{R}/L_{\textrm{loc}}$
for potentials $V_{\delta,sp}$ and $V_{\delta,D}$ the Dirac-delta approximations of the potentials $V_{sp}$ and $V_D$ respectively. We focus on two cases where the disorder of the pottential is $0.04E_{\sigma_R}$ or $0.5E_{\sigma_R}$. When generating potentials $V_{\delta,sp}$ and $V_{\delta,D}$ we assume that it is the variance of the potential being approximated that is equal to one of the above values.

For the case of low variance of the potentials $0.04E_{\sigma_R}$ the inverse localization lengths is shown in Fig.~\ref{fig:loclength2}a). For small $k\sigma_R$ the inverse localization lengths in all four cases are similar. This is because, for shallow disorder the series expansion given by \eqref{eq:locCorrel} holds and $\sigma_R/L_{\textrm{loc}}$ is determined by the variance of the potential that closely match. 

For  $k\sigma_R$ near 1, we observe "kinks" in the dependence of $\sigma/L_{\textrm{loc}}$ on $k\sigma_R$. In case of the speckle potential this is followed by a sudden drop of $\sigma_R / L_{\textrm{loc}}$. This is in a stark contrast to the Dirac-delta approximation of the speckle potential $V_{\delta,{sp}}$ (and $V_{\delta,D}$). This is not surprising as the speckle potential is smooth and Fourier transform of its correlation function has finite support. We saw in previous sections that for  dark state potentials the $\tilde{C}_2$ contained arbitrarily high nonzero Fourier components explaining why $\sigma_R / L_{\textrm{loc}}$  for $V_{D}$ and $V_{\delta,D}$ are closer than for $V_{sp}$ and $V_{\delta,sp}$. The agreement of $\sigma_R / L_{\textrm{loc}}$ for  $V_{\delta,D}$ of $V_{D}$ may be regarded as at most qualitative for $k\sigma_R$. Still the Dirac-delta approximation of $V_D$ reproduces the fine details of the dependence of $\sigma_R/L_{\textrm{loc}}$ on $k\sigma_R$ such as the kink at $k\sigma_R=1$.

For the deeper disorder with potential variance of $0.5E_{\sigma_R}$, we see in Fig.~\ref{fig:loclength2}b) that the $\sigma_R / L_{\textrm{loc}}$  for $V_{D}$ and $V_{\delta,D}$ nearly match. This is because the dark state potential consists now of well-defined narrow peaks, which are well approximated by discrete Dirac-deleta peaks of $V_{\delta,D}$. The difference shows up for only very high momenta, beginning from $k\sigma_R \approx 3.5$.

For both shallow and deeper disorder potential, one can reach the conclusion that the Dirac delta $V_{\delta,D}$ potential is a valid approximation for the low-energy part of the spectrum of Hamiltonian of a particle moving in the dark state potential $V_D$ (only qualitative for a shallow disorder). This is in contrast to the localization in a speckle field which cannot be described by a Kronig-Penney-like model.

\section{Tight-binding description of movement in random comb potential}
\label{section:RandomComb}

In this Section we discuss localization in the dark state potenial $V_D$ for the configuration presented in Section~\ref{subsec:The-dark-state-running-lambda} for  $\Omega_1(x)=\tilde{\Omega}_1\sin(k_1 x),$ $\Omega_2(x)=\tilde{\Omega}_2S_2(x)+\tilde{\Omega}_2^0$, when the potential consists narrow peaks separated by $a=\pi/k_1$.
The low energy dynamics in such a potential is captured by a Dirac-delta approximation $V_{D,\delta}$, Eq.~\eqref{eq:kp} with $V_n$ given by \eqref{eqn:effHeight} and $x_n= n a$. Localization in such a lattice has been previously intensively studied \cite{Izrailev2012}. Following that review, we consider the Schr\"odinger equation in the following form:

\begin{align}
\left[-\frac{\hbar^2}{2m} \frac{d^2}{dx^2}\!\!+\!\!\!\!\!\sum\limits_{n=-\infty}^{\infty} \!\!\!E_r(\bar{V}\!+\!\delta V_n)\delta(k_1 x\!-\!k_1x_n\!)\!\right]\!\psi(x)\!=&&\nonumber\\
=\!\frac{\hbar^2 q^2}{2m}\psi(x),&&
\label{eq:izkp}
\end{align}
where $\bar{V} + \delta V_n = V_n$, $\langle \delta V_n \rangle_{n\in \mathbb{Z}} =0$, $\sigma^2=\langle (\delta V_n)^2 \rangle_{n\in \mathbb{Z}}  \ll \bar{V}^2 $.
Under above assumptions the inverse localization length, $L_{\textrm{loc}}$ is:

\begin{figure}
\includegraphics[width=8.6cm]{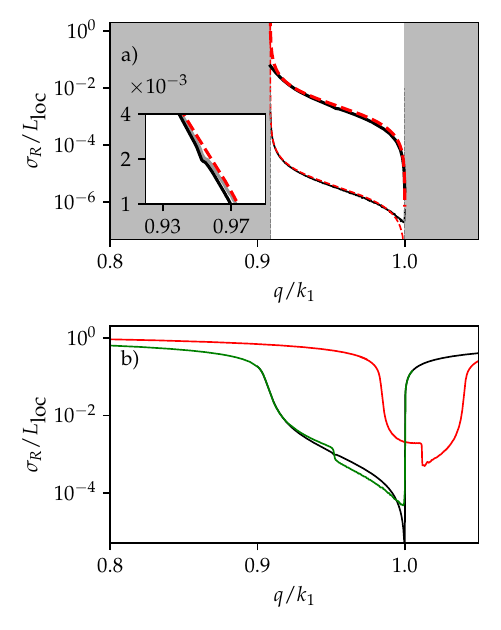}

\caption{The inverse localization length in  Kronig-Penney-like potentials with compositional disorder. Panel (a) random Dirac delta peaks, thick lines: $\epsilon_+=0.15, \epsilon_-=0.1$, thin lines: $\epsilon_+=0.126, \epsilon_-=0.124$;  solid black lines: numerical calculation of $\sigma_R/L_{\textrm{loc}}$ for sample length $L=16\cdot 10^6 /k_1$; red dashed lines show analytical,  Eq.~\eqref{eq:invlocDelta}. The gray areas denote $q$'s not in applicability interval of this equation. Panel (b)  green line: $\sigma_R/L_{\textrm{loc}}$ for $\epsilon_+=0.15, \epsilon_-=0.1$ for $V_{\delta,D}$, $\delta=0.0005$. Black: random uncorrelated Dirac delta scatterers $\epsilon_+=0.1444,$ $\epsilon_-=0.1087$, red: localization length in the dark state potential $V_D$ for $\Omega_1(x)=\tilde{\Omega}_1\sin(k_1x), \Omega_2(x)=\tilde{\Omega}_2S_2(x)+\tilde{\Omega}_2^0$, $\epsilon_+=0.15, \epsilon_-=0.1$, green: $V_{D,\delta}$ approximation of $V_D(x)$ given by the red curves. The $k_1$ is set such that $k_1\sigma_R=1$ and $R/f=1/3$.
}
\label{fig:plotBandLoc}

\end{figure}

\begin{equation}
\frac{a}{L_{\textrm{loc}}}=\frac{1}{8} \frac{k_1^2\sin qa}{q^2\sin^2ka} \sum\limits_{l=-\infty}^{\infty}\!\!\langle \delta V_n \delta V_{n+l}\rangle_{n\in \mathbb{Z}}\cos(2 ka l).
\label{eq:invlocDelta}
\end{equation}
The wavevector $k$ is obtained from
\begin{equation}
\cos(ka)=\cos(qa)+\frac{\bar{V }k_1}{2 E_r q}\sin(qa).
\label{eq:blochK}
\end{equation}
for those $q$ that correspond to the band in case of $\delta V_n=0$. The equation \eqref{eq:invlocDelta} is valid only for those $q$'s and it cannot be applied in the forbidden bands.  There, in presence of disorder, the density of states is exponentially suppressed \cite{Lifshits1,Lifshits2,Lifshits3}, but it is  non-zero. The localization for those energies  can be addressed numerically. 
{Additionally, the analytic expression is not expected to hold near the bottom and the top of the band. Another limitation follows from the details of derivation of Eq.~\eqref{eq:invlocDelta} (see \cite{Izrailev2012}): the latter does not yield an anomaly in the localization length at the band centre. It predicts a smooth dependence of $L_{\textrm{loc}}$.  }

\paragraph{Random uncorrelated disorder.-} We first consider \eqref{eq:izkp} with random, uncorrelated $\delta V_n$.  The exact values $V_n =\frac{\pi}{2\epsilon}$ are based on random value of $\epsilon$ with uniform distribution in $[0.124,0.126]$ (weak disorder case) and in $[0.1,0.15]$ (strong disorder case). The mean potential heights are $\bar{V}=12.56$ and $12.73$ respectively. The intervals of applicability of Eq.~\eqref{eq:invlocDelta} are  $q/k_1\in[0.9096, 1]$ and $q/k_1 \in [0.9085, 1]$. There \eqref{eq:blochK} can be solved for Bloch momentum.

In Figure~\ref{fig:plotBandLoc}a) we compare the localization length given by Eq.~\eqref{eq:invlocDelta} to the numerically determined $L_{\textrm{loc}}$ as the function of $q/k_1$ for $q$ within the regions of validity marked with vertical gray dashed lines \footnote{the line corresponding to $q/k_1=0.9096$ is very close to $q/k_1=0.9085$ and was thus omitted}.  We find the quantitatively good agreement between the localization length obtained from Eq.~\eqref{eq:invlocDelta} (red lines) and from numerics (black lines). This is true for both weak disorder (thin lines) and strong disorder cases  {(thick lines)}.
{The discrepancies appear near band edges where the analytical expression for the inverse correlation length diverges  or equals to zero}
 The singularity present in the dependence of $L_{\textrm{loc}}$ on $q/k_1$ determined numerically [see inset in Fig.~\ref{fig:plotBandLoc}a)] is absent in the analytic expression, Eq.~\eqref{eq:invlocDelta}. 

\paragraph{Dark state potential.-} We now consider the full dark state potential for $\Omega_1(x)=\tilde{\Omega}_1\sin(k_1 x),$ $\Omega_2(x)=\tilde{\Omega}_2S_2(x)+\tilde{\Omega}_2^0$. 
For comparison, we again consider Hamiltonian Eq.~\eqref{eq:invlocDelta} with $\delta V_n$ based on $V_{\delta,D}$ approximation, Eq.~\eqref{eqn:effPos} and \eqref{eqn:effHeight}.
For the parameters considered in this section the potential $V_D$ consists of isolated, well-defined peaks. This allows us to use $\delta=0.0005$, much smaller than $\delta=0.25$ used in Section~\ref{subsec:Dirac-Delta-approximation}.
We focus on the strong disorder case where for the dark state potential $\epsilon_+=0.15$ and $\epsilon_-=0.1$. The majority of peaks of the dark state potential is between $44.4 E_r$ and $100E_r$. When the integral \eqref{eqn:effHeight} is calculated, this gives the $V_{\delta,D}$ consisting of Dirac deltas with $\bar{V} = 12.51$. We can also find parameters of the Hamiltonian with uncorrelated $\delta V_n$'s that will have the same $\bar{V}$, the amplitude of the disorder is matched by requiring that the standard deviation of $\delta V_n$'s is the same. Fulfilling those two requirements results in parameters $\epsilon_+=0.1444$ and $\epsilon_-=0.1087$ for the random $\delta V_n$ case.

We now compare the localization length determined numerically for the dark state potential $V_D(x)$ and for the $V_{\delta,D}$ Dirac-delta approximation. In Fig.~\ref{fig:plotBandLoc}b) we show $a/L_{\textrm{loc}}$ as the function of $q/k_1$ (respectively red and green lines). In both cases  the inverse localization length shows a dip for the values of $\hbar^2 q^2 / {2m}$ that can be traced back to the conduction band in the case of no disorder. The visible difference in the location of this region in $q/k_1$ is due to a finite width of  potential peaks. 

One also can observe that the singularity near the band centre is very pronounced. It is much larger than in the random uncorrelated disorder case (see black curve in Fig.~\ref{fig:plotBandLoc}b)). This occurs for both $V_D(x)$ potential and its Dirac-delta approximation $V_{\delta,D}(x)$. This is in contrast to the model with random and uncorrelated $\delta V_n$'s with  $\bar V_n$ and the standard deviation {of $V_n$ matching that of $V_{\delta,D}(x)$. }
{We find that correlations between $V_n$'s  in $V_{\delta,D}$ and in the $V_D(x)$ potential enhance the amplitude of the band centre anomaly. 
This is a known possible effect of disorder correlation \cite{Titov2005,Izrailev2012}.
}

Moreover for the actual dark state potential $V_D(x)$ the localization length $L_{\textrm{loc}}$ does not monotonically increase with $q/k_1$. The maximal $L_{\textrm{loc}}$ is reached near the anomaly band center, not at the top of the band like in $V_{\delta,D}(x)$.

\section{Conclusions and outlooks}

We have shown the construction of the potential for ultracold atoms using
a three-level atomic system. The potential applies to the ultracold atoms populating
the dark state. The potentials
consist of narrow pseudo-random peaks, with randomness driven by the
speckle laser field. We have contrasted the properties of the dark
state potential against the off-resonant optical lattice potential
given by the speckle field. 

We have found substantially enhanced localization in the dark state
potential, especially for high kinetic energy of the particle. This
is explained by a slow decay of the two-point correlation function
in the Fourier space, a manifestation of the non-linearity of the dark state
potential. This is rooted in different mechanism of generation of
the dark state potential than that of the speckle potential which
is due to far off-resonant AC-Stark process. 

Our findings indicate that the potential generation via a dark state of a three level 
system enhances the resolution of the speckle potential and
preserves its randomness properties. This can be further extended
by replacing speckle fields generating Rabi frequency $\Omega_{1}$
with a laser standing wave. It
leads to a completely different class of potentials that consist of
tall, pseudorandom potential peaks implementing e.g a Kronig-Penney
model structural disorder akin to \citep{Sanchez1994,Izrailev2001}.

\begin{acknowledgments}
M.\L . and J.Z. acknowledge support from National Science Centre (Poland)
through grants No. 2019/35/B/ST2/00838 and 2019/35/B/ST2/00034, respectively.
The research has been supported by a grant from the Priority Research Area (DigiWorld) under the Strategic Programme Excellence Initiative at Jagiellonian University. 
No part of this work was written by the artificial intelligence.
\end{acknowledgments}

\bibliographystyle{apsrev4-2}
%\bibliography{biblio}

\end{document}